# Spin-state transition, magnetism and local crystal structure in $Eu_{1-x}Ca_xCoO_{3-\delta}$

A.N. Vasiliev[1], T.M. Vasilchikova[1], O.S. Volkova[1], A.A. Kamenev[1], A.R. Kaul[1], T.G. Kuzmova[1], D.M. Tsymbarenko[1], K.A. Lomachenko[2], A.V. Soldatov[2], S.V. Streltsov[3,4], J.-Y. Lin[5], C.N. Kao[5], J.M. Chen[6], M. Abdel-Hafiez[7], A.U.B. Wolter[7], and R. Klingeler[8]

*[1]Low Temperature Physics and Superconductivity Department,a*
*M.V. Lomonosov Moscow State University, 119991 Moscow, Russia*
*[2]Southern Federal University, 344090 Rostov-on-Don, Russia*
*[3]Institute of Metal Physics, 620990 Ekaterinburg, Russia*
*[4]Ural Federal University, 620002 Ekaterinburg, Russia*
*[5] Institute of Physics, National Chiao-Tung University, 30076 Hsinchu, Taiwan*
*[6] National Synchrotron Radiation Research Center, Hsinchu 30076, Taiwan*
*[7]Leibniz Institute for Solid State and Materials Research, IFW Dresden, 01171 Dresden, Germany*
*[8]Kirchhoff Institute for Physics, University of Heidelberg, 69120 Heidelberg,Germany*

The doping series $Eu_{1-x}Ca_xCoO_{3-\delta}$ provides a rather peculiar way to study the spin-state transition in cobalt-based complex oxides since partial substitution of $Eu^{3+}$ ions by $Ca^{2+}$ ions does not increase the mean valence state of cobalt but is accompanied by appearance of oxygen vacancies in the ratio $\delta \sim x/2$. In the parent compound $EuCoO_3$, the low spin (LS) – high spin (HS) transition takes place at temperatures so high that the chemical decomposition prevents its direct observation. The substitution of $Eu^{3+}$ for $Ca^{2+}$ in this system shifts the LS-HS transition to lower temperatures. The energy gap $\Delta$ associated with this transition in octahedrally-coordinated $Co^{3+}$ ions changes from 1940 K in $EuCoO_3$ to 1540 K in $Eu_{0.9}Ca_{0.1}CoO_{2.95}$ and 1050 K in $Eu_{0.8}Ca_{0.2}CoO_{2.9}$. Besides, each $O^{2-}$ vacancy reduces the local coordination of two neighboring $Co^{3+}$ ions from octahedral to pyramidal thereby locally creating magnetically active sites which couple into dimers. These dimers at low temperatures form another gapped magnetic system with very different energy scale, $D \sim 3$ K, on the background of intrinsically non-magnetic lattice of octahedrally-coordinated low-spin $Co^{3+}$ ions.

## Introduction

Among the transition metal oxides the cobalt-based ones exhibit the richest variety of valence and spin states.[1] In $RCoO_3$ perovskites (R is a rare-earth) the cobalt ions are presumed to be in the 3+ state retaining 6 electrons in the d-shell.[2] Depending on the ratio of local crystal field splitting $\Delta_{CF}$ between $t_{2g}$ and $e_g$ levels and the Hund's coupling $J_H$, the $Co^{3+}$ ions may exhibit either low-spin ($t_{2g}^6e_g^0$), intermediate-spin ($t_{2g}^5e_g^1$), or high-spin ($t_{2g}^4e_g^2$) states. The crystal field splitting is defined by the oxygen environment of the $Co^{3+}$ ions. In case of a Hund's energy predominance and octahedral coordination, the high-spin state (S = 2) is Jahn-Teller active due to partial filling of the $t_{2g}$ shell which yields local distortions of the ligand cages. At the increase of $\Delta_{CF}$, the stable low-spin state (S = 0) can be realized. The possibility of the intermediate-spin state (S = 1) is still under debate and might be derived from multiplet calculations when including spin-orbit coupling and the Jahn-Teller instability in both the $t_{2g}$ and $e_g$ segments of the d-shell.[3,4] Due to the subtle balance of $\Delta_{CF}$ and $J_H$ in $LaCoO_3$, the crossover between LS, IS and/or HS states can be studied experimentally since modest variation of temperature induces changes in the spin state between the low-spin state at low temperature and a magnetic IS or HS state. In $LaCoO_3$, the magnetization passes through a broad maximum at about 100 K and follows the Curie law at elevated temperatures.[5] At even higher temperatures, the magnetization increases again and shows temperature independent behavior inherent for itinerant magnetic systems.[6]

This metal-insulator transition at about 500 K is usually ascribed to the melting of the orbitally ordered state.

The substitution of $La^{3+}$ ions by any other $R^{3+}$ ions of smaller radius results in chemical pressure which shifts both the spin state transition and the metal-insulator transition to higher temperatures[7-10]. The effects are very different for both transitions, however. For instance, in $EuCoO_3$ the spin state transition takes place at about 2000 K while the temperature of the metal-insulator transition temperature is about 600 K. This strongly different effect of chemical pressure on both transitions emphasizes the different origin of the spin state and the metal-insulator transitions.[11]

The substitution of $R^{3+}$ ions by alkaline earth $AE^{2+}$ ions may cause the appearance of tetravalent $Co^{4+}$ ions. This should drastically change the magnetic and electronic properties in $R_{1-x}AE_xCoO_3$ series of compounds due to presumed double exchange interactions between $Co^{3+}$ and $Co^{4+}$ ions. To be specific, ferromagnetism and increased conductivity may occur while, at commensurate doping levels, metal-insulator transitions can occur due to charge ordering phenomena.[12,13] This scenario however demands exact oxygen stoichiometry in the material. The preparation of $R_{1-x}AE_xCoO_3$ compounds at ambient oxygen pressure usually results in oxygen deficient samples $R_{1-x}AE_xCoO_{3-\delta}$. In this case, the formation of $Co^{4+}$ ions is energetically unfavorable and only $Co^{3+}$ ions are present. Since $AE^{2+}$ doping is associated with chemical pressure significant effects on the spin-gap and the metal insulator transition are to be expected. In addition, the removal of $O^{2-}$ ions from the crystal structure should change the local environment of both rare-earth and transition metals in the oxygen deficient samples. While this effect is less relevant for the rare-earths with their deeply hidden f-shells, the reduction in transition metal coordination may change the spin state of electrons in the d-shells.

In the present work, X-ray diffraction (XRD) and X-ray absorption near edge structure (XANES) spectra were taken on $Eu_{1-x}Ca_xCoO_{3-\delta}$ series of compounds in order to evaluate the trends in the crystal lattice parameters and local environment of the transition metal Co upon substitution of trivalent Eu for divalent Ca and oxygen deficiency. The specific heat and magnetization in $Eu_{1-x}Ca_xCoO_{3-\delta}$ were studied in order to reveal the associated effects on the thermodynamic properties at low temperatures and to evaluate trends in the spin-state transitions upon doping at high temperatures. The choice of $Eu^{3+}$ ions among the rare earths metals was motivated by the fact that these ions exhibit a half-filled f-shell and hence a vanishing total magnetic moment, i.e. the rather small contribution of the transition metal subsystem is not masked. The $Ca^{2+}$ ions among alkaline earths metals were chosen because the difference in ionic radii with the $Eu^{3+}$ ions was expected to shift the spin-state transitions to lower temperatures.

### The synthesis and XRD spectra

Polycrystalline samples of $Eu_{1-x}Ca_xCoO_{3-\delta}$ with $0 \le x \le 0.4$ were prepared by the chemical homogenization (paper synthesis) method.[14] At first, non-concentrated water solutions of metal nitrates $Eu(NO_3)_3$, $Ca(NO_3)_2$ and $Co(NO_3)_2$ of 99.95 % purity were prepared. The exact concentration of the dissolved chemicals was determined by gravimetry and in case of Co-based solutions by means of complexonometric titration. The weighted amounts of metal nitrates solutions were mixed in stoichiometric ratios and calculated mixtures of nitrates were dropped on the ash-free filters. The filters were dried at about 80°C and the procedure of the solutions dropping was performed repeatedly. Then, the filters were burned and the remaining ash was thoroughly grinded. It was annealed at 600°C for 4 hours to remove the carbon. The powder obtained was pressed into pellets and sintered at 800°C in oxygen atmosphere for 30 hours. Finally, the samples were slowly cooled down to room temperature by switching off the furnace. The oxygen content in the samples studied was determined by means of iodometric titration with 2% precision. The chemical analysis shows that at $x > 0.2$ the decomposition of the samples occurs, so that for further investigations the samples with $x = 0$, 0.1 and 0.2 were selected.

The samples of $Eu_{1-x}Ca_xCoO_{3-\delta}$ ($x = 0$, 0.1, and 0.2) were analyzed at room temperature by powder X-ray diffraction in symmetrical reflection geometry using Rigaku SmartLab (Cu $K_\alpha$ radiation, λ=1.54187Å, graphite monochromator). All detectable peaks were indexed in the orthorhombic space group P*nma*, as given in Table 1.

The full-profile Rietveld refinement was performed using JANA2006 package.[15] The representative XRD pattern for $EuCoO_3$ is shown in Fig. 1. The patterns were fitted with ten-term Legendre polynomial background and four-term pseudo-Voight shape function. The zero-shift and sample roughness were also refined. For all crystallographically independent atoms the atomic positions and isotropic thermal parameters were refined. The oxygen thermal parameters were constrained to be identical. The refinement of oxygen occupancies demonstrates the large correlations between occupancies, thermal parameters and sample roughness. So, unfortunately, it could not be refined against the present XRD data.

The structure of the $Eu_{1-x}Ca_xCoO_{3-\delta}$ (x = 0, 0.1, and 0.2) consists the eight-fold coordinated Eu (x=0) or Eu/Ca (x=0.1, 0.2) atoms and six-fold coordinated Co atoms in distorted octahedral environment. There are two non-equivalent positions for the oxygen ions O1 and O2 which constitute distorted $CoO_6$ octahedron connected via shared corners. The four O2 atoms are arranged in the equatorial plane while two O1 atoms are situated at the octahedron vertices. Rather small changes in the unit cell parameters, the unit cell volume remains practically constant, reflect the fact that the withdrawal of the $O^{2-}$ ions is practically compensated by substitution of $Eu^{3+}$ ions by larger $Ca^{2+}$ ions. As follows from the Table 2, the structural changes in the $Eu_{1-x}Ca_xCoO_{3-\delta}$ (x = 0, 0.1, and 0.2) series of compounds are not univocal. The various Eu-O and Co-O distances change monotonically with the increasing x but in different directions. Thus, the average Eu-O and Co-O distances remain constant equal to 2.439Å and 1.930Å respectively.

Partial substitution of Eu by Ca leads to the increasing distortion of $CoO_6$ octahedron. Moreover the Co – Co distance shrinkage along the *b* axis, i.e. distance between Co ions bound by O1 ions, with the simultaneous increase of Co – Co distance in the *ac*-plane. This could be explained by withdrawal of apical oxygen ions O1 with the appearance of $Ca^{2+}$ ions.

The removal of any oxygen changes the local environment of two neighboring $Co^{3+}$ ions from octahedron to pyramid. As will be shown below, these changes in local environment of the $Co^{3+}$ ions lead to changes in their spin state. To be specific, in pyramidal coordination the splitting of $t_{2g}$ and $e_g$ levels results in the formation of either the intermediate S = 1 state[16] or the high S = 2 spin-state.[17] In both cases, magnetically active $Co^{3+}$ ions arise which may interact and form magnetic dimers.

**The XANES spectra**

The XANES spectra of $Eu_{1-x}Ca_xCoO_{3-\delta}$ (x = 0, 0.1 and 0.2) taken at liquid nitrogen temperature are shown in Fig. 2a. At partial substitution of Eu for Ca and accompanying oxygen deficiency the positions of main $L_{2,3}$ peaks remain virtually unchanged, but new features appear, i.e. the pre-edge peak A strengthens, the C peak diminishes and strong peak D arises on the right shoulders of both $L_2$ and $L_3$ edges. These extra D peaks at 784.5eV and 799 eV increase in amplitude with x in $Eu_{1-x}Ca_xCoO_{3-\delta}$ series of compounds.

Theoretical simulations of Co $L_3$-edge XANES spectra in $Eu_{1-x}CoCa_xO_{3-\delta}$ were carried out using real-space Green's function formalism implemented in self-consistent FEFF 8.4 code.[18,19] Full multiple scattering approach with Hedin-Lundqvist exchange-correlation potential was applied.[20] Muffin-tin approximation was employed for the shape of potential. This method is well-developed and has been proven to give useful structural information regarding substitutions, vacancies and defects in various types of materials.[21,22]

The model used for calculations was the spherical fragment of $EuCoO_3$ lattice (18 Å in diameter), built around the absorbing cobalt atom. In this structure nearest neighbors of cobalt are six oxygens occupying the O1 and O2 non-equivalent crystallographic sites. Although all O2 atoms are crystallographically equivalent, with respect to each particular cobalt atom there are O2a and O2b atoms with slightly different Co-O distances, as shown in the inset to Fig. 2b. Since doping of $EuCoO_3$ by calcium causes the creation of oxygen vacancies, firstly the spectra of pure and doped models with different number of oxygen vacancies were calculated to estimate which factor (Eu-Ca substitution or creation of vacancies) has more influence on XANES spectra. In order to simulate the effect of doping, eight europium atoms, nearest to absorbing cobalt, were replaced by calcium. It was found, that the only major consequence of Eu-Ca substitution is the increase of pre-edge intensity (peak A). Apart from that, spectra of pure and doped material with the same configuration of vacancies are very similar. It means that decrease of C feature and appearance of D peak are mainly caused by the creation of oxygen vacancies, which occurs as a result of the doping, but not by the Eu-Ca substitution itself.

Even the appearance of a single oxygen vacancy around the absorbing atom introduces significant changes to the theoretical spectrum. Intensity of the C peak decreases with respect to the spectrum without vacancies, but this decrease is not sufficient to turn the D shoulder into a separate peak. Intensity and position of pre-edge peak A also differ from the experiment. To reproduce changes in the experimental spectrum multiple vacancy model was used. Final comparison between experimental and theoretical spectra for pure and Ca-doped $EuCoO_3$ is shown in Fig. 2b. Theoretical spectra are broadened in order to match the experimental resolution. Positions and relative intensities of peaks are in good agreement with the experiment for both pure and doped samples. The only significant mismatch is too low intensity of D shoulder in the theoretical spectrum of doped model.

To check the origins of the present mismatch and the possible effect of chemicals contamination on these extra features, a specially designed experiment was conducted. XANES spectra were compared as shown in Fig. 3 for pure rare-earth cobaltates $EuCoO_3$, $SmCoO_3$, the Ca-doped cobaltates $Eu_{0.8}Ca_{0.2}CoO_{2.9}$ and $Sm_{0.8}Ca_{0.2}CoO_{2.9}$, and the mechanical mixture of $SmCoO_3$ and $0.4Ca(NO_3)_2$ which contains significant excess of Ca as compared to $Eu_{0.8}Ca_{0.2}CoO_{2.9}$ and $Sm_{0.8}Ca_{0.2}CoO_{2.9}$ samples. Indeed, the XANES spectra of $SmCoO_3$ + $0.4Ca(NO_3)_2$ sample contains additional features on the right shoulders of both $L_3$ and $L_2$ peaks, as shown in Fig. 3. Tentatively, this can be attributed to the presence of small quantities of Ba in the sample namely in $Ca(NO_3)_2$, so that the barium $M_{4,5}$ edge is visible in the experimental spectra and overlaps with D feature. From the data in Fig. 3, nominal purity $Ca(NO_3)_2$ chemical provides about half of the observed feature D intensity. Nevertheless, additional features seen on the right shoulders of both $L_3$ and $L_2$ peaks are significantly more pronounced in magnitude. In theory the effect of Ba contamination is not taken into account; therefore, these calculated peaks are less pronounced compared with the observed ones.

**Specific heat**

The temperature dependences of specific heat C were measured in the range 2 – 30 K in "Quantum Design" Physical Properties Measurements System PPMS-9T by the relaxation technique. The C(T) curves obtained for $Eu_{1-x}Ca_xCoO_{3-\delta}$ series of compounds are shown in Fig. 4. In this temperature range the C(T) dependence in the parent compound $EuCoO_3$ can be well approximated by just the cubic term $\beta T^3$ but a tiny Schottky – type anomaly is observed at low temperatures, nevertheless. This anomaly becomes significantly more pronounced in the $Eu_{0.9}Ca_{0.1}CoO_{2.95}$ and $Eu_{0.8}Ca_{0.2}CoO_{2.90}$ samples. Therefore, at low temperatures the specific heat can be approximated by the sum $C = \beta T^3 + C_{Sch}$.

As follows from XRD and XANES analyses, the partial substitution of trivalent Eu by divalent Ca is accompanied by the appearance of apical oxygen vacancies. Taking into account that the removal of $O^{2-}$ ions reduces the coordination number of the neighboring $Co^{3+}$ ions, one can assume that the position of cobalt with respect to oxygens will change in a way that the cation ion will move within the pyramid. In this case the spin state of $Co^{3+}$ may change from the non-magnetic LS (S = 0) to a magnetic either intermediate IS (S = 1) or high spin HS (S = 2) state. Two neighboring ions in pyramid coordination may weakly interact with each other forming the magnetic dimers. Assuming the magnetic nature of the Schottky – type anomaly one can approximate its shape by the formula for S = 1

$$C = 6nR\left(\frac{J}{kT}\right)^2 \frac{\exp\left(-\frac{2J}{kT}\right)\left(1+15\exp\left(-\frac{4J}{kT}\right)+20\exp\left(-\frac{6J}{kT}\right)\right)}{\left(1+3\exp\left(-\frac{2J}{kT}\right)+5\exp\left(-\frac{6J}{kT}\right)\right)^2} \quad (1)$$

or similar but more cumbersome expression for S = 2, where n is the concentration of dimer entities, R = 8.314 J/mol K is the universal gas constant, and J is the exchange interaction parameter within the dimer. The fits of available data by Eq. 1 are shown in the Inset to Fig. 4. Within the limits of accuracy, fitting by means of Eq. 2 describes the data within similar quality, i.e. the data do not discriminate between the S=2 and the S=1 case. The estimation of the phonons contribution gives $\beta = (1.7\pm0.1)\times10^{-4}$ which corresponds to the Debye temperature of about 400 K. The gap value deduced from the Schottky anomaly position was found to be D = 2J = 3.2 ± 0.1 K, while for $Eu_{0.9}Ca_{0.1}CoO_{3-\delta}$ and $Eu_{0.8}Ca_{0.2}CoO_{3-\delta}$ samples the concentrations n were found to be 0.052 and 0.098, correspondingly. These numbers (n = x/2) are in good agreement with expected 0.05 and 0.1 for x = 0.1 and 0.2.

**Magnetization**

The temperature dependences of the magnetization in the $Eu_{1-x}Ca_xCoO_{3-\delta}$ family of compounds were measured by means of a VSM-SQUID magnetometer (Quantum Design) in the temperature range 2 – 390 K in both zero-field-cooled (ZFC) and field-cooled (FC) regimes at B = 0.1 T. In case of the parent compound $EuCoO_3$, ZFC and FC curves practically coincide in the whole temperature range measured. Therefore, hereinafter the curves measured in ZFC regime are discussed only.

The temperature dependences of the static magnetic susceptibility $\chi = M/B$ for the family of $Eu_{1-x}Ca_xCoO_{3-\delta}$ samples taken at B = 0.1 T are shown in Fig. 5. Firstly, the magnetic response of the undoped material $EuCoO_3$ is discussed. The data are in a good qualitative agreement to previous studies on single- and polycrystalline materials.[11] As shown in Fig. 5, below about T = 250 K, the magnetization is nearly perfectly described by the van Vleck magnetism of the $Eu^{3+}$ ions and a Curie-Weiss-like divergence of the magnetic susceptibility. Temperature independent diamagnetic and $Co^{3+}$ van Vleck contributions are very small in comparison. The Curie-Weiss-like contribution implies a negligible number of paramagnetic sites in the material, i.e. of either HS or IS Co-ions or $Eu^{2+}$ ions, cf. with tiny Schottky-type anomaly found in the specific heat of the nominally pure parent compound $EuCoO_3$. At higher temperature there is an additional contribution to the magnetization which we ascribe to the temperature-induced evolution of $Co^{3+}$ ions in the HS or IS state. A quantitative analysis of this feature will be given below.

Upon Ca-doping, several effects show up in the data in Fig. 5: (i) there is a strong increase of the Curie-Weiss-like low-temperature response; (ii) the pronounced shoulder at intermediate temperature which signals a dominant $Eu^{3+}$ van Vleck contribution to the susceptibility in this temperature range is suppressed; (iii) the susceptibility at high temperatures increases. Qualitatively, these features suggest the increase of paramagnetic sites at low-temperatures, the decreasing relevance of the $Eu^{3+}$ van Vleck magnetism and the decrease of the magnetic excitation gap associated with the $Co^{3+}$ HS- or IS-state formation.

In the following, the various contributions to the susceptibility will be discussed in more details. In accordance with the analysis of our XRD and XANES data, we assume that the substitution of $Eu^{3+}$ ions by $Ca^{2+}$ ions is accompanied by the formation of $Co^{3+}$ ion pairs in pyramidal coordination, i.e. of pairs of paramagnetic $Co^{3+}$ sites forming antiferromagnetically coupled magnetic dimers. Generally, the expression for the magnetic susceptibility in the $Eu_{1-x}Ca_xCoO_{3-\delta}$ family of compounds can hence be written as a sum of virtually independent terms:

$$\chi = \chi_0 + \chi_{Eu} + \chi_{Co\text{-}pyr} + \chi_{Co\text{-}oct} \quad (2)$$

where $\chi_0$ is the temperature-independent diamagnetic and Co van Vleck contribution, $\chi_{Eu}$ is the van Vleck contribution of the $Eu^{3+}$ ions, $\chi_{Co-pyr}$ is the contribution of magnetically active $Co^{3+}$ ions in pyramidal coordination, and $\chi_{Co-oct}$ is the $Co^{3+}$ ions contribution dependent on their spin state in octahedral coordination.

The procedure of separation of various terms to the magnetic susceptibility could be illustrated for the parent compound $EuCoO_3$, as follows. As discussed above, at low temperatures, the main contributions to the magnetization are due to the $Eu^{3+}$ van Vleck term $\chi_{Eu}$ and a Curie-Weiss-like term due to a small amount of paramagnetic sites. The van Vleck term can be written as

$$\chi_{Eu} = \frac{N_A \mu_B^2}{3k_B T} \frac{(\frac{24}{\gamma}) + (13.5 - 1.5\gamma)e^{-\gamma} + (67.5 - 2.5\gamma)e^{-3\gamma} + ...}{1 + 3e^{-\gamma} + 5e^{-3\gamma} + ...} \tag{3}$$

The only fitting parameter in this formula $\gamma = \lambda/k_BT$ is a constant of the spin-orbital interaction, i.e. the multiplet width.[23] This contribution is almost temperature independent below *ca.* 100 K and decreases proportional to 1/T at high temperatures. The best fit of the experimental data at intermediate temperatures could be obtained with the multiplet width $\lambda = 311$ $cm^{-1}$ (~ 450 K). This value is in good correspondence with the values of this parameter in other Eu-based oxides.[24,25] In Figs. 5 and 6, the results of fitting of the van Vleck contribution $\chi_{Eu}$ are shown as dashed lines. Note, that the summation of the Pascal's constants for $EuCoO_3$, i.e. $\chi_0 = -6.6\times10^{-5}$ emu/mol,[26] and a $Co^{3+}$ van Vleck contribution are negligible as compared to other terms.

As mentioned above, our data suggest that all relevant contributions to the magnetic susceptibility change for the samples with $x \neq 0$. In such a situation, when the magnetic susceptibility $\chi$ comprises contributions differently dependent on temperature, it is more convenient to represent the experimental data as temperature dependences of the product $\chi\cdot T$, as shown in Fig. 6. In this case, hyperbolic terms will be temperature independent.

In the oxygen-deficient samples $Eu_{1-x}Ca_xCoO_{3-\delta}$, i.e. for x = 0.1 and 0.2, the quantity $\chi\cdot T$ demonstrates doping-induced changes both at low and high temperatures while at intermediate temperatures there are only small effects observed. However, as will be shown in more details below we ascribe the latter to an incidental compensation of decreasing $Eu^{3+}$ van Vleck magnetism by the increasing relevance of the other terms.

At low temperature, upon doping a clear maximum evolves in $\chi\cdot T$. Indeed, such a feature can be straightforwardly explained in the frame of our analysis by the formation of magnetic dimers. In this scenario, the magnetic susceptibility probes the dimer excitation gap, i.e. the size of the gap is given by the peak maximum, the associated susceptibility changes measure the number of dimers, and the peak width provides information on the distribution of gap energies. Our data hence indicate that the number of dimers increases upon Ca-doping and that the magnetic interaction within the dimers is only small, i.e. of the order of few Kelvin degrees, which agrees both with the specific heat measurements and the band structure calculation results presented in the next section.

At high temperatures, the divergence of curves, shown in Fig. 6, is due the contribution of $Co^{3+}$ ions in octahedral coordination. At elevated temperatures, these $Co^{3+}$ ions gradually transform from the non-magnetic low spin state (S = 0) to a magnetic either intermediate spin (S = 1) or high spin (S = 2) state. The temperature dependence of the magnetic susceptibility $\chi_{Co-oct}$ can be hence approximated by the expression for two-level system:[11]

$$\chi(T) = \frac{N_A g^2 \mu_B^2 S(S+1)}{3k_B T} \frac{\nu(2S+1)e^{-\Delta/T}}{1 + \nu(2S+1)e^{-\Delta/T}} \tag{4}$$

where $\nu = 6$ is the degeneracy of the intermediate spin state and $\nu = 3$ is the degeneracy of the high spin state in $Co^{3+}$ ions, the gap $\Delta$ is the splitting between the low-spin LS and excited either

IS or HS state. An estimation of the gap $\Delta$ can be made directly from this expression by analyzing the high temperature magnetic susceptibility.

In Fig. 6 the product $\chi \cdot T$ in the parent compound $EuCoO_3$ is compared to the $Eu^{3+}$ van Vleck contribution. The difference of these curves, shown in the lower inset of Fig. 6, can be approximated by the magnetic susceptibility given by Eq. (5) multiplied by temperature. Fitting the data yields the energy gap $\Delta$ between the low-spin and intermediate-spin state in $EuCoO_3$ of $\Delta$ = 1940 K. This value nicely agrees to a previous analysis of single crystal data which yielded $\Delta$ > 1900 K.[11] Upon doping, the magnetization at high temperatures increases which already qualitatively indicates that the energy gap $\Delta$ becomes smaller. Indeed, calculations similar to that for the parent compound allow establishing the concentration dependence of the energy gap $\Delta$ in the $Eu_{1-x}Ca_xCoO_{3-\delta}$ family of compounds, as shown in Fig. 7. Evidently, the gap decreases rapidly in the range doping $0 < x < 0.2$ reaching $\Delta$ = 1540 K in $Eu_{0.9}Ca_{0.1}CoO_{2.95}$ and $\Delta$ = 1050 K in $Eu_{0.8}Ca_{0.2}CoO_{2.9}$. It extrapolates through polynomial to zero at about x = 0.365.

Our data hence show that the energy gap $\Delta$ between the low-spin and intermediate-spin state strongly decreases upon doping. This behavior agrees to what is expected for changing the chemical pressure in $EuCoO_3$ by partial substitution of the europium ions (the ionic radius of $Eu^{3+}$ is 120 pm) for larger calcium ions (the ionic radius of $Ca^{3+}$ is 126 pm). A similar tendency is observed when Eu is replaced by smaller La as well as in hydrostatic external pressure experiments on $LaCoO_3$.[27]

### Intra-dimer magnetic exchange

In order to estimate the size of the spin gap in the dimers we performed calculations of the exchange parameters J in the Heisenberg model within the density functional theory using Green's function formalism.[28] We used the linearized muffin-tin orbitals (LMTO) method[29] and the LDA+U approximation.[30] The on-site Coulomb repulsion U and intra-atomic exchange parameter $J_H$ were chosen to be 7.8 eV and 0.99 eV respectively.[5] The Eu-4f states were considered as pseudocore.[31] The integration in the Brillouin zone was performed over 144 k-points.

We started with the $EuCoO_3$ crystal structure and constructed the supercell with 8 Co atoms. According to our XRD data apical oxygen ions are primary removed with the doping. In the calculations we followed the same scenario and removed these oxygens to simulate the dimers (formed by two pyramidal Co) directed along c-axis. The lack of the charge due to the oxygen withdrawal was compensated by the corresponding shift of the Fermi level. Since our aim is to calculate the magnetic properties at lowest temperatures we assumed that all octahedral Co ions are in the low-spin state, while pyramidal Co were allowed to be magnetic.

The exchange parameter J within the dimer was found to be 3.1 K. This is in agreement with the specific heat and magnetic susceptibility measurements. Since in the high-spin state stabilized in the pyramidal $Co^{3+}$ ions[16] $e_g$ orbitals are half-filled they participate in the antiferromagnetic superexchange interaction within the dimer. In the present case, when the dimers are along c-axis, $Co^{3+}$ $z^2$-orbitals are active. However, since in the high spin state both $e_g$ orbitals are actually half-filled the same mechanism (but with $x^2$-$y^2$ orbital) is expected for the planar order of the dimers Co, even if such will be the case.

### Conclusion

The synthesis of $Eu_{1-x}Ca_xCoO_{3-\delta}$ at ambient pressure yields a series of compounds with unchanged Co valence states since the partial substitution of $Eu^{3+}$ ions for $Ca^{2+}$ ions is accompanied by oxygen deficiencies, as observed in XRD and XANES measurements. This family of compounds hence provides an alternative way to change the local structure and to study the spin state transitions. Here, we have presented XANES, XRD, specific heat, and magnetization data in order to elucidate the various contributions to the magnetic response in this system where only $Co^{3+}$ ions are present. In octahedral coordination, these ions exhibit a non-magnetic S = 0 ground state at low temperatures and experience a spin state transition at increasing temperatures. The energy gap Δ associated with LS-HS transition in octahedrally-coordinated $Co^{3+}$ ions changes from 1940 K in $EuCoO_3$ to 1540 K in $Eu_{0.9}Ca_{0.1}CoO_{2.95}$ and 1050 K in $Eu_{0.8}Ca_{0.2}CoO_{2.9}$. The analysis of the product $\chi$·T allows revealing the major trends in behavior of the systems experiencing the transition between different spin states.

The removal of the oxygen ion from the perovskite crystal lattice changes the local environment of two neighboring cobalt ions from octahedral to pyramidal. The $Co^{3+}$ ions in pyramidal coordination are in a magnetic (either S = 1 or S = 2) ground state at low temperatures. The weak interaction between these spins results in the formation of magnetic dimers which show up in the low temperature magnetic susceptibility $\chi$ and specific heat C through appearance of Schottky – type anomaly with D ~ 3 K. The intra-dimer exchange calculation performed within the density functional theory confirms this estimation.

**Acknowledgements**

We thank K.I. Kugel, D.I. Khomskii, B. Büchner, and Z. Hu for fruitful discussions. This work was supported through RFBR Grant No. 10-02-00021, 10-02-96011, 11-02-91335 and by the DFG via 486 RUS 113/982/0-1.

Fig. 1. Room temperature powder XRD pattern (black circles) of $EuCoO_3$ in the 2θ range 10 – 90°, Rietveld refinement fit (solid red line), difference profile (lower solid green line), and positions of Bragg peaks (vertical bars). In the inset the primitive cell of $EuCoO_3$ is shown. The Eu ions are indicated by large gray circles, the Co ions both bare and in octahedral oxygen environment are represented by blue circles and two positions of oxygen ions are shown by small red (O1) and black (O2) circles.

Fig. 2. The Co $L_{2,3}$-edge XANES spectra in $Eu_{1-x}Ca_xCoO_{3-\delta}$ series of compounds taken at liquid nitrogen temperature (a); the comparison of experimental and simulated XANES spectra in pure $EuCoO_3$ and Ca-doped compounds (b). The spectra were shifted along y-axis for clarity.

Fig. 3. The Co $L_{2,3}$-edge XANES spectra in $EuCoO_3$, $SmCoO_3$, $Eu_{0.8}Ca_{0.2}CoO_{2.9}$ $Sm_{0.8}Ca_{0.2}CoO_{2.9}$ and in the mechanical mixture of $SmCoO_3$ and $0.4Ca(NO_3)_2$ powders.

Fig. 4. The temperature dependences of specific heat C in $Eu_{1-x}Ca_xCoO_{3-\delta}$ series of-compounds. Inset: the approximation of the specific heat in Ca-doped compounds by sum of cubic term $\beta T^3$ (not shown) and Schottky-type anomaly $C_{Sch}$ (dashed lines).

Fig. 5. Temperature dependence of the magnetic susceptibility $\chi$ of $Eu_{1-x}Ca_xCoO_{3-\delta}$. The dashed line represents the $Eu^{3+}$ van Vleck contribution for the parent compound $EuCoO_3$.

Fig. 6. The temperature dependences of the product $\chi \cdot T$ for the $Eu_{1-x}Ca_xCoO_{3-\delta}$ family of compounds. Lower inset represents the difference between the data for x = 0 and the $Eu^{3+}$ van Vleck contribution (black data and dashed line, respectively, in main plot). The upper inset highlights the low temperature behavior.

Fig. 7. The concentration dependence of the gap Δ between low spin and high spin states in the $Eu_{1-x}Ca_xCoO_{3-\delta}$ family of compounds. The dashed line is a polynomial extrapolation of the experimental data.

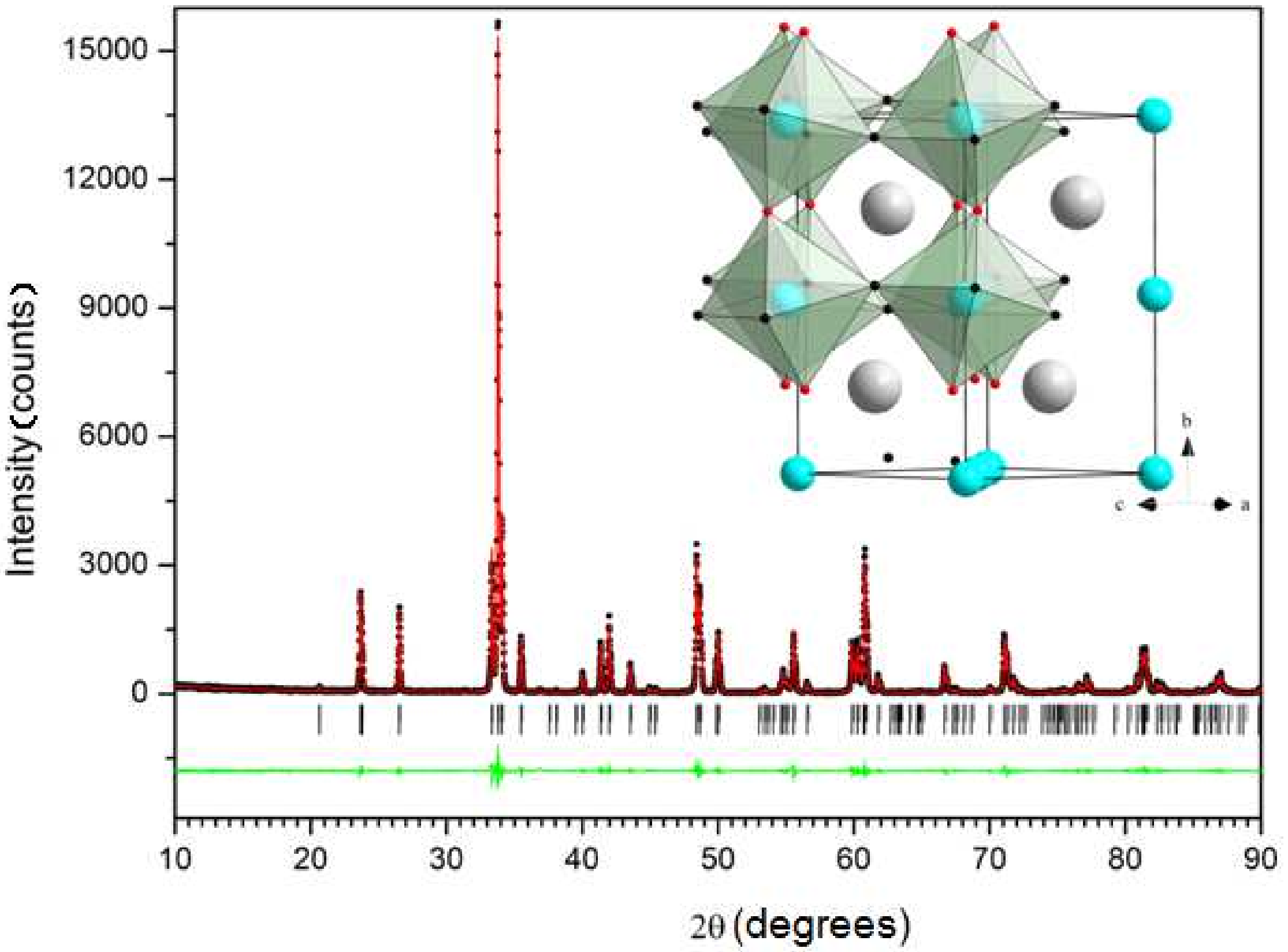

Fig. 1.

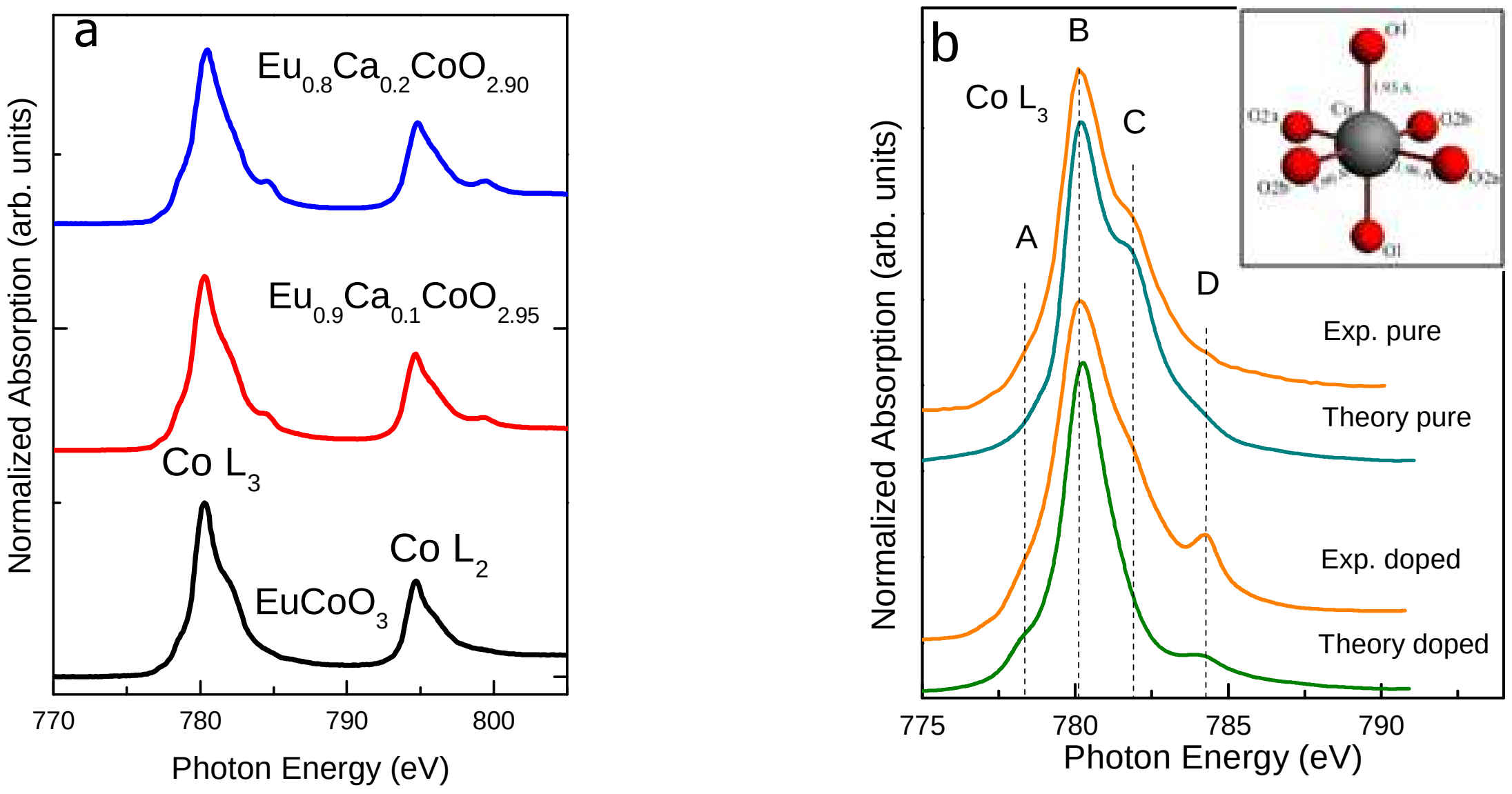

Fig.2.

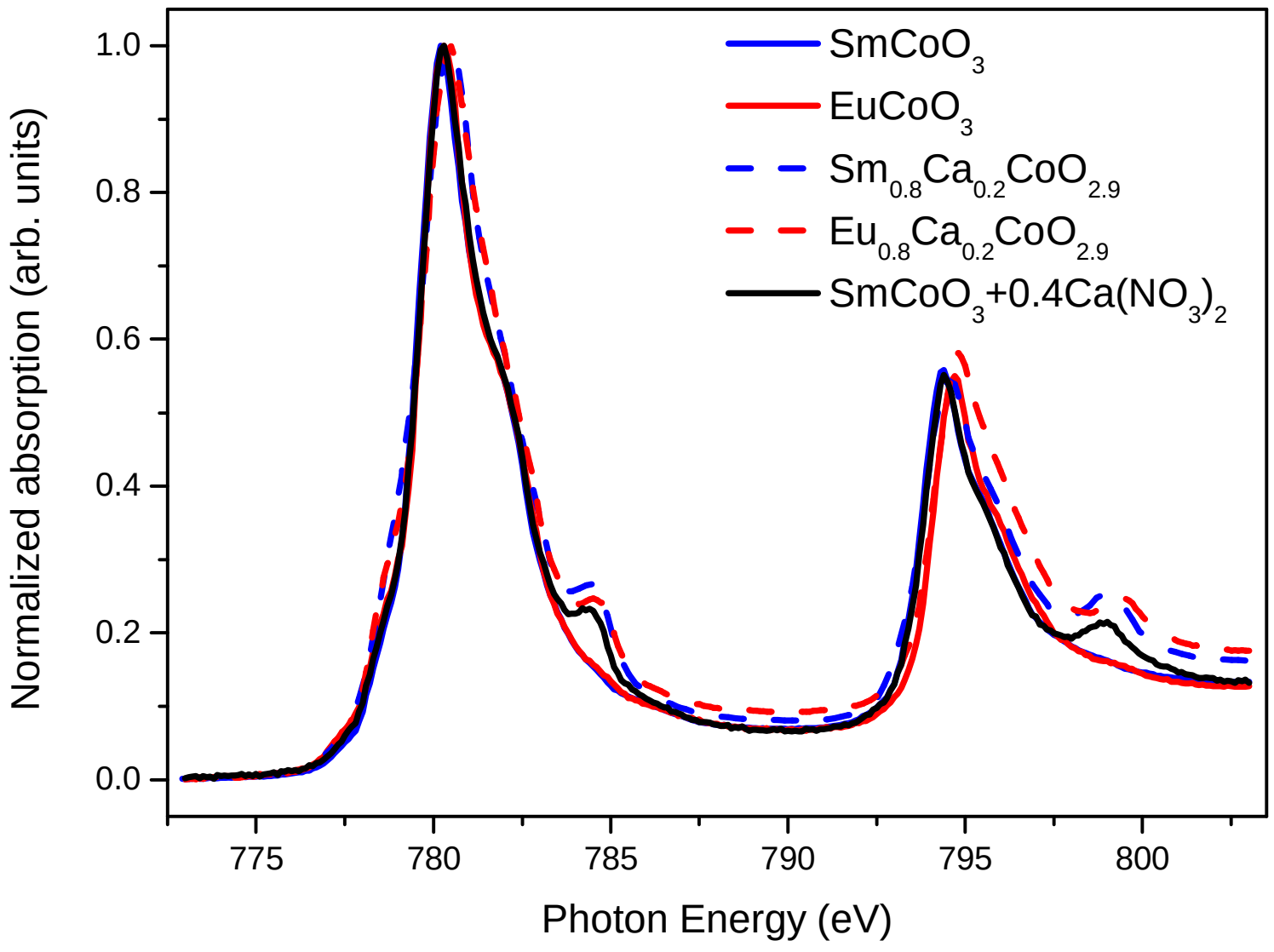

Fig. 3

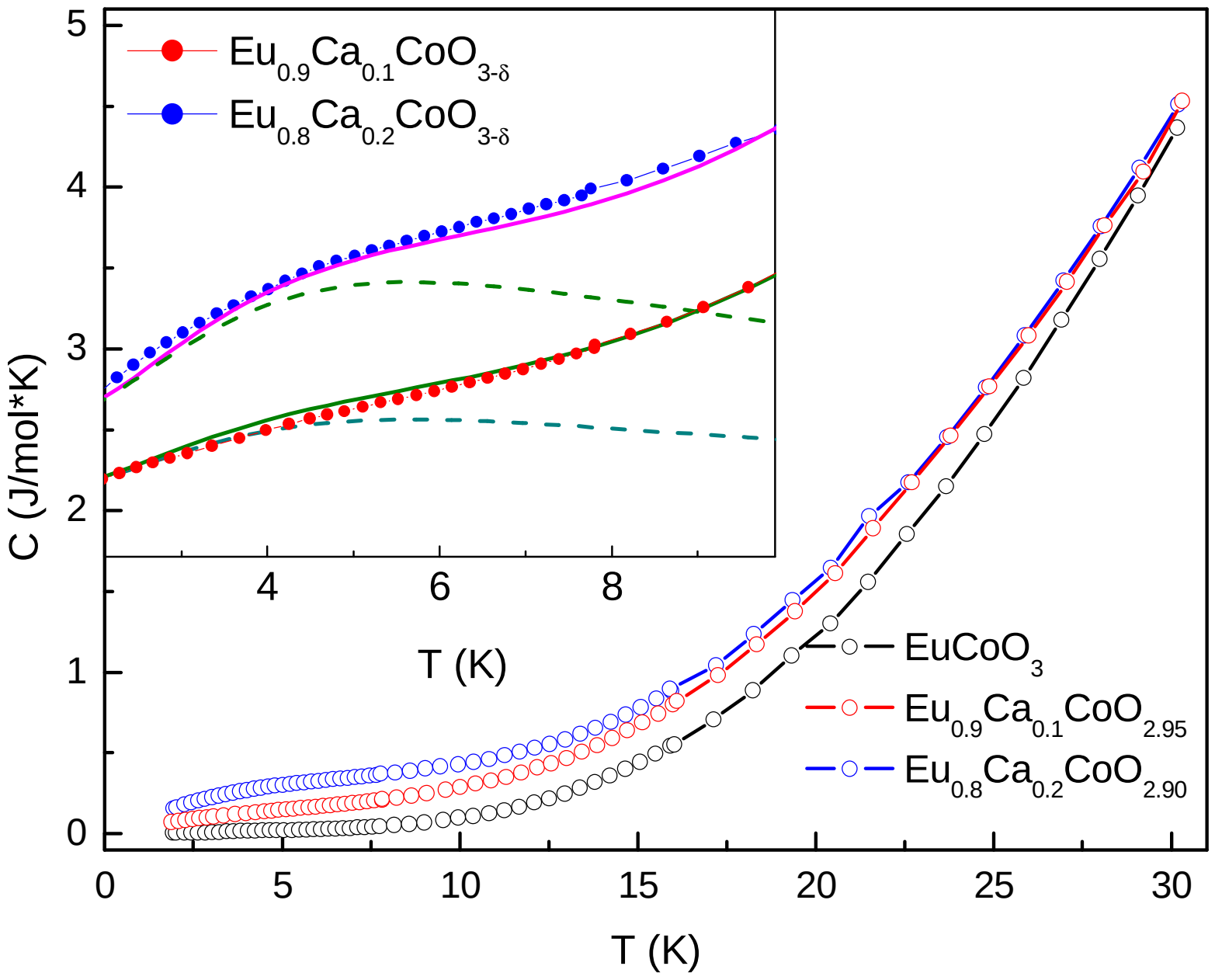

Fig. 4.

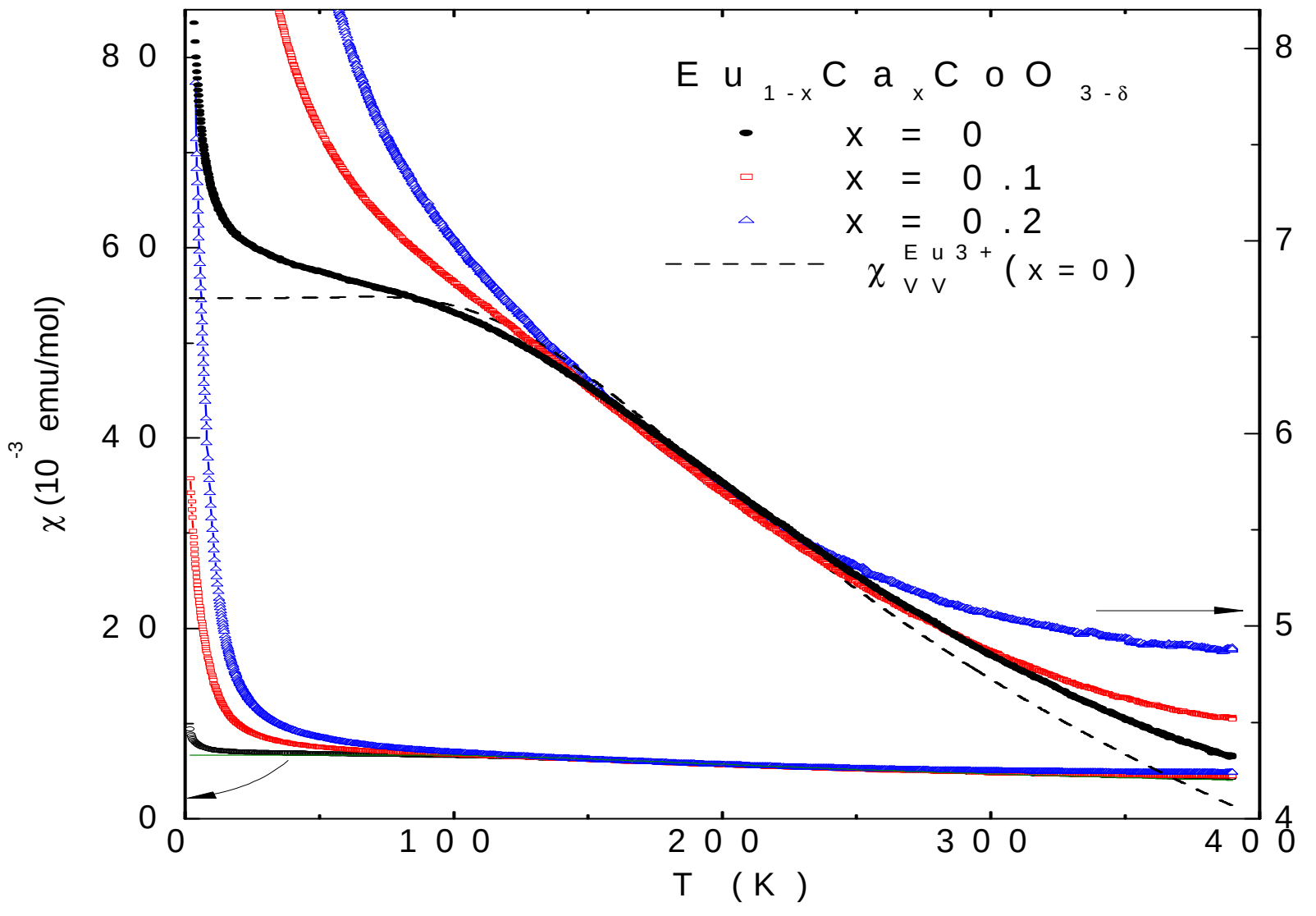

Fig. 5.

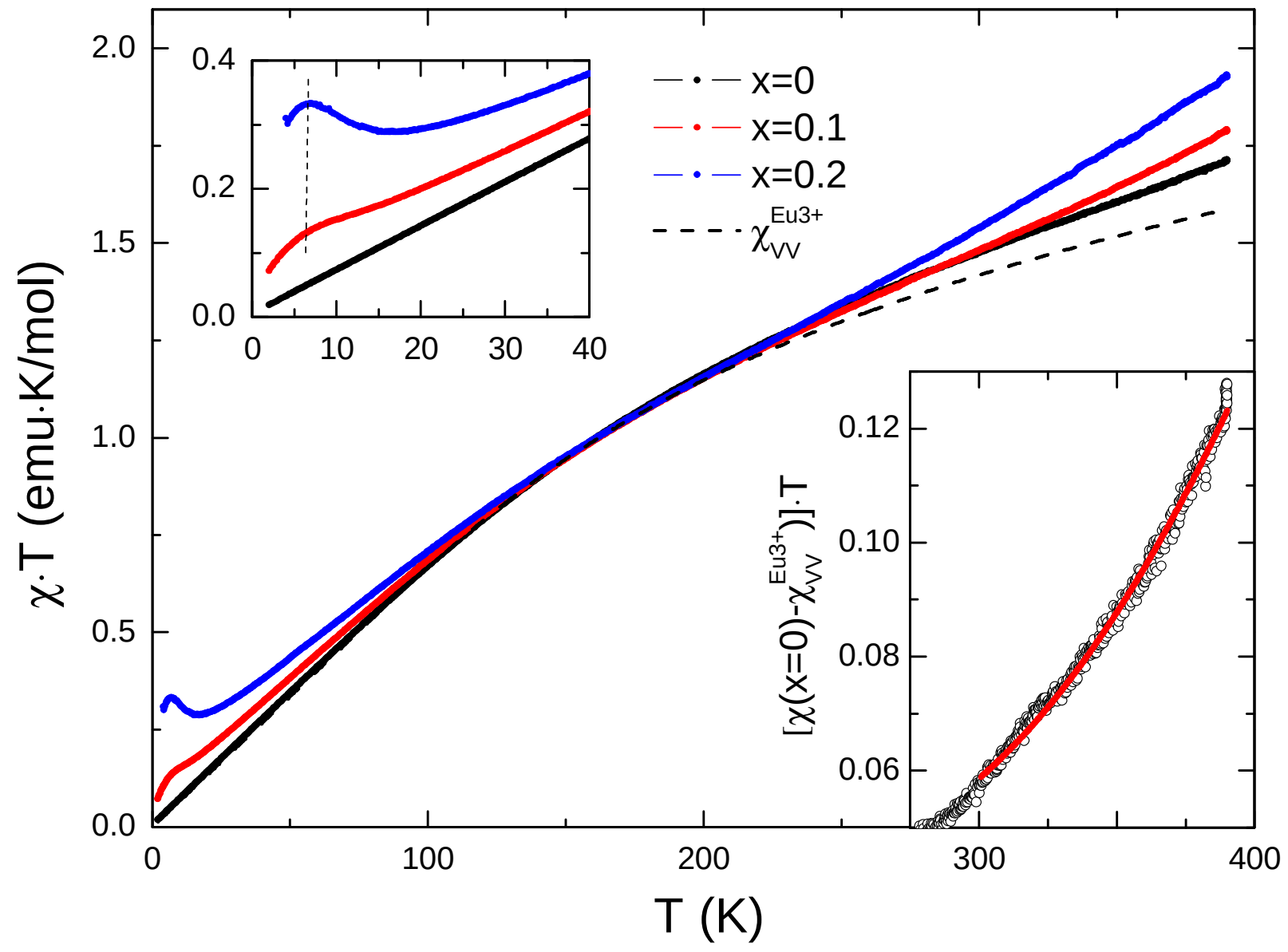

Fig. 6.

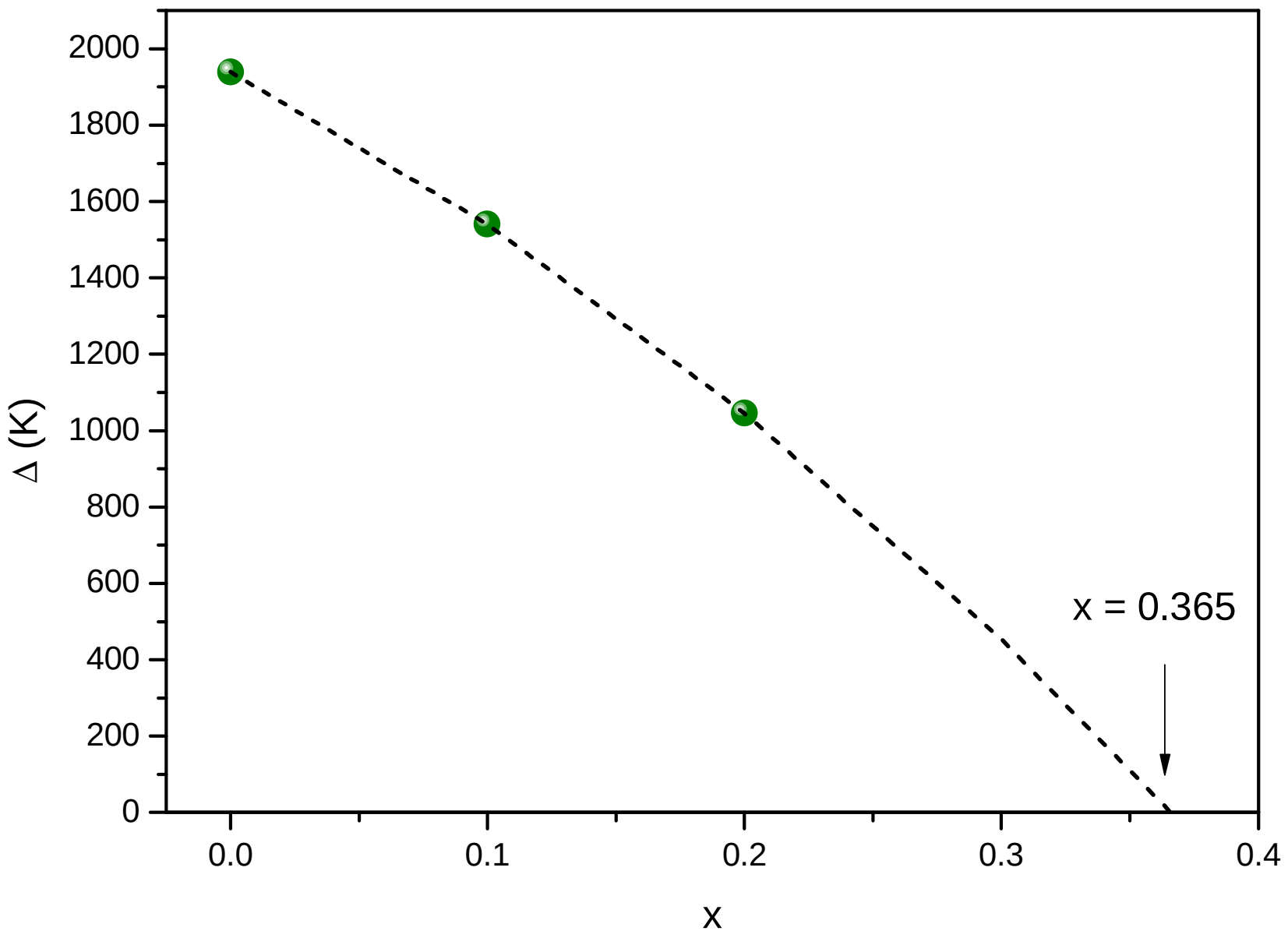

Fig. 7.

Table 1. Crystallographic parameters of the orthorhombic $Eu_{1-x}Ca_xCoO_{3-□}$ system at room temperature (space grooup Pnma). Listed are the lattice parameters a, b and c and the quality-of-fit parameters $R_1/\omega R_2$, $Rp_1/\omega Rp_2$, and Goodness-of-Fit (GOF).

| Compound | a (Å) | b (Å) | c (Å) | V, $Å^3$ | Nobs/Ntotal/ Nstr.param/ Ntotal param | $R_1$/ $\omega R_2$ | Rp/ ωRp | GOF |
|---|---|---|---|---|---|---|---|---|
| $EuCoO_3$ | 5.3715(2) | 7.4785(2) | 5.2572(2) | 211.19(2) | 90/94/10/28 | 0.0160/ 0.0202 | 0.0843/ 0.1256 | 1.69 |
| $Eu_{0.9}Ca_{0.1}CoO_{2.95}$ | 5.3653(2) | 7.4763(2) | 5.2632(2) | 211.12(2) | 92/94/9/28 | 0.0147/ 0.0192 | 0.0690/ 0.1074 | 1.58 |
| $Eu_{0.8}Ca_{0.2}CoO_{2.90}$ | 5.3609(2) | 7.4749(3) | 5.2691(2) | 211.15(2) | 91/93/9/26 | 0.0166/ 0.0231 | 0.0710/ 0.1103 | 1.72 |

Table 2. Selected interatomic distances in $Eu_{1-x}Ca_xCoO_{3-□}$ series of compounds at room temperature.

| Parameter | $EuCoO_3$ | $Eu_{0.9}Ca_{0.1}CoO_{2.95}$ | $Eu_{0.8}Ca_{0.2}CoO_{2.90}$ |
|---|---|---|---|
| Eu1 – $O2^{iii}$, Eu1 – $O2^{iv}$ | 2.320(9) | 2.314(8) | 2.311(11) |
| Eu1-O2, Eu1-$O2^{v}$ | 2.506(9) | 2.516(8) | 2.528(11) |
| Eu1-$O2^{i}$, Eu1-$O2^{vi}$ | 2.605(9) | 2.604(8) | 2.600(11) |
| Eu1 – $O1^{i}$ | 2.294(11) | 2.285(11) | 2.264(14) |
| Eu1 – $O1^{ii}$ | 2.359(11) | 2.364(10) | 2.374(12) |
| Co1-O2, Co1-$O2^{vii}$ | 1.926(9) | 1.922(8) | 1.912(11) |
| Co1-$O2^{i}$, Co1-$O2^{viii}$ | 1.942(9) | 1.947(8) | 1.956(10) |
| Co1-O1, Co1-$O1^{ii}$ | 1.921(3) | 1.921(2) | 1.925(3) |
| Co-$Co^{i}$ (in *ac* plane) | 3.7580(2) | 3.7579(2) | 3.7584(1) |
| Co-$Co^{ii}$ (along *b* axis) | 3.7393(2) | 3.7381(2) | 3.7374(2) |

Symmetry codes: (i) −*x*+1/2, −*y*, *z*+1/2; (ii) −*x*, *y*+1/2, −*z*; (iii) *x*+1/2, −*y*+1/2, −*z*−1/2; (iv) *x*+1/2, *y*, −*z*−1/2; (v) *x*, −*y*+1/2, *z*; (vi) −*x*+1/2, *y*+1/2, *z*+1/2; (vii) −*x*, −*y*, −*z*; (viii) *x*−1/2, *y*, −*z*−1/2.